\documentclass[12pt]{article}

\catcode`\@=11

\@addtoreset{equation}{section}
\def\theequation{\thesection\arabic{equation}}

\def\@normalsize{\@setsize\normalsize{15pt}\xiipt\@xiipt
\abovedisplayskip 14pt plus3pt minus3pt%
\belowdisplayskip \abovedisplayskip
\abovedisplayshortskip  \z@ plus3pt%
\belowdisplayshortskip  7pt plus3.5pt minus0pt}
\def\small{\@setsize\small{13.6pt}\xipt\@xipt
\abovedisplayskip 13pt plus3pt minus3pt%
\belowdisplayskip \abovedisplayskip
\abovedisplayshortskip  \z@ plus3pt%
\belowdisplayshortskip  7pt plus3.5pt minus0pt
\def\@listi{\parsep 4.5pt plus 2pt minus 1pt
            \itemsep \parsep
            \topsep 9pt plus 3pt minus 3pt}}

\def\underline#1{\relax\ifmmode\@@underline#1\else
        $\@@underline{\hbox{#1}}$\relax\fi}
\@twosidetrue \relax

\catcode`@=12

\catcode`\@=11

\def\section{\@startsection{section}{1}{\z@}{3.5ex plus 1ex minus
   .2ex}{2.3ex plus .2ex}{\large\bf}}
\def\thesection{\arabic{section}.}

\def\ps@headings{\def\@oddfoot{}\def\@evenfoot{}
\def\@oddhead{\hbox{}\hfill
        \makebox[.5\textwidth]{\raggedright\ignorespaces --\thepage{}--
        \hfill }}
\def\@evenhead{\@oddhead}
\def\subsectionmark##1{\markboth{##1}{}}
}

\ps@headings

\catcode`\@=12

\relax

\def\figcap{\section*{Figure Captions\markboth
        {FIGURECAPTIONS}{FIGURECAPTIONS}}\list
        {Fig. \arabic{enumi}:\hfill}{\settowidth\labelwidth{Fig. 999:}
        \leftmargin\labelwidth
        \advance\leftmargin\labelsep\usecounter{enumi}}}
 \relax
\def\tablecap{\section*{Table Captions\markboth
        {TABLECAPTIONS}{TABLECAPTIONS}}\list
        {Table \arabic{enumi}:\hfill}{\settowidth\labelwidth{Table 999:}
        \leftmargin\labelwidth
        \advance\leftmargin\labelsep\usecounter{enumi}}}
 \relax
\def\reflist{\section*{References\markboth
        {REFLIST}{REFLIST}}\list
        {[\arabic{enumi}]\hfill}{\settowidth\labelwidth{[999]}
        \leftmargin\labelwidth
        \advance\leftmargin\labelsep\usecounter{enumi}}}
 \relax

\catcode`\@=11

\def\marginnote#1{}
\newcount\hour
\newcount\minute
\newtoks\amorpm
\hour=\time\divide\hour by60 \minute=\time{\multiply\hour by60
\global\advance\minute by- \hour}
\edef\standardtime{{\ifnum\hour<12 \global\amorpm={am}%
    \else\global\amorpm={pm}\advance\hour by-12 \fi
    \ifnum\hour=0 \hour=12 \fi
    \number\hour:\ifnum\minute<100\fi\number\minute\the\amorpm}}
\edef\militarytime{\number\hour:\ifnum\minute<100\fi\number\minute}
\def\draftlabel#1{{\@bsphack\if@filesw {\let\thepage\relax
  \xdef\@gtempa{\write\@auxout{\string
    \newlabel{#1}{{\@currentlabel}{\thepage}}}}}\@gtempa
    \if@nobreak \ifvmode\nobreak\fi\fi\fi\@esphack}
     \gdef\@eqnlabel{#1}}
\def\@eqnlabel{}
\def\@vacuum{}
\def\draftmarginnote#1{\marginpar{\raggedright\scriptsize\tt#1}}
\def\draft{\oddsidemargin -.5truein
        \def\@oddfoot{\sl preliminary draft \hfil
        \rm\thepage\hfil\sl\today\quad\militarytime}
        \let\@evenfoot\@oddfoot \overfullrule 3pt
        \let\label=\draftlabel
        \let\marginnote=\draftmarginnote

\def\@eqnnum{(\theequation)\rlap{\kern\marginparsep\tt\@eqnlabel}%
\global\let\@eqnlabel\@vacuum}  }
\def\preprint{\twocolumn\sloppy\flushbottom\parindent 1em
        \leftmargini 2em\leftmarginv .5em\leftmarginvi .5em
        \oddsidemargin -.5in    \evensidemargin -.5in
        \columnsep 15mm \footheight 0pt
        \textwidth 250mmin      \topmargin  -.4in
        \headheight 12pt \topskip .4in
        \textheight 175mm
        \footskip 0pt

\def\@oddhead{\thepage\hfil\addtocounter{page}{1}\thepage}
        \let\@evenhead\@oddhead \def\@oddfoot{} \def\@evenfoot{}
}
\def\titlepage{\@restonecolfalse\if@twocolumn\@restonecoltrue\onecolumn
     \else \newpage \fi \thispagestyle{empty}\c@page\z@
        \def\thefootnote{\fnsymbol{footnote}} }
\def\endtitlepage{\if@restonecol\twocolumn \else  \fi
        \def\thefootnote{\arabic{footnote}}
        \setcounter{footnote}{0}}  
\catcode`@=12 \relax

\def\ps@headings{\def\@oddfoot{}\def\@evenfoot{}
\def\@oddhead{\hbox{}\hfill
        \makebox[.5\textwidth]{\raggedright\ignorespaces --\thepage{}--
        \hfill }}
\def\@evenhead{\@oddhead}
\def\subsectionmark##1{\markboth{##1}{}}
}

\ps@headings

\relax

\usepackage{graphicx}

\begin{document}

\begin{titlepage}
\begin{flushright}

NTUA--99--2000

hep-th/0011051 \\

\end{flushright}

\begin{centering}
\vspace{.41in}
{\large {\bf Brane Inflation from Mirage Cosmology.}}\\

\vspace{.2in}

{\bf E.~Papantonopoulos}$^{a}$  \\

\vspace{.2in}

 National Technical University of Athens, Physics
Department, Zografou Campus,\\ GR 157 80, Athens, Greece. \\

\vspace{0.5in}

{\bf Abstract} \\

\end{centering}

\vspace{.3in}
We study the cosmological evolution of a D3-brane universe in a
type 0 string background. We follow the brane-universe along the
radial coordinate of the background and we calculate the energy
density which is induced on the brane because of its motion in the
bulk. We find that for some typical values of the parameters the
brane-universe has an inflationary phase.

\vspace{1.5in}
\begin{flushleft}

Talk given at IX Marcel Grossmann Meeting, Rome, July 2000

$^{a}$ e-mail address:lpapa@central.ntua.gr

\end{flushleft}

\end{titlepage}

\section{Introduction}

There has been much recent interest in the idea that our universe
 may be a brane embedded in some higher dimensional space
 \cite{Reg}. It has been shown that the hierarchy problem can be
 solved if the higher dimensional Planck scale is low and the
 extra dimensions large \cite{Dim}. Randall and Sundrum \cite{Rand}
 proposed a solution of the hierarchy problem
 without the need for large extra dimensions but instead through
 curved five-dimensional spacetime $AdS_{5}$ that generates an
 exponential suppression of scales.

 This idea of a brane-universe can  naturally be applied to string
 theory. In this context, the Standard Model gauge bosons as well as
 charged matter arise as fluctuations of the D-branes. The universe
 is living on a collection of coincident branes, while gravity and
 other universal interactions is living in the bulk space
 \cite{Pol}.

 This new concept of brane-universe naturally leads to a new
 approach to cosmology. Any cosmological evolution like inflation
 has to take place on the brane while gravity acts globally on
 the whole space. In the literature there are a lot of cosmological
 models which study the cosmological evolution of our universe. In
 most of these models the spacetime is five-dimensional, where
 the fifth dimension is the radial dimension of an $AdS_{5}$
 space. The effective Einstein equations on the brane are then solved
 taking under consideration the matter on the brane  \cite{Cosm}-\cite{Keh}.

 Another approach to cosmological evolution of our brane-universe
 is to consider the motion of the brane in  higher
 dimensional spacetimes. In \cite{Cha} the motion of a domain wall
 (brane) in such a space  was studied. The Israel matching conditions
 were used to relate the bulk to the domain wall (brane) metric, and
 some interesting cosmological solutions were found. In
 \cite{Keh} a universe three-brane is considered in motion
 in ten-dimensional space in the presence of a gravitational field
 of other branes. It was shown that this motion in ambient space
 induces cosmological expansion (or contraction) on our universe,
 simulating various kinds of matter.

 In this direction we have studied \cite{Papa}, the motion of a three-brane
 in the background of type 0 string theory . It was shown
 that the motion of the brane on this specific background,
 with constant values of dilaton and tachyon fields,
 induces a cosmological evolution which for some range of the
 parameters has an inflationary phase. In \cite{Kim},
 using similar technics  the
 cosmological evolution of the three-brane in the background of
 type IIB string theory was considered.

 Type 0 string theories \cite{Tset}
 are interesting because of their connection
 \cite{Typ0} to four-dimensional
 $SU(N)$ gauge theory. Asymptotic solutions of the theory were
 constructed in \cite{Tset,Minah}. At large radial coordinate
 the tachyon is constant and one
 finds a metric of the form $AdS_{5} \times S^{5}$ with vanishing
 coupling which was interpreted as a UV fixed point. The solution
 exhibits a logarithmic running in qualitative agreement with the
 asymptotic freedom property of the field theory. At small radial
 coordinate the tachyon vanishes and one finds again a solution of the
 form $AdS_{5} \times S^{5}$ with infinite coupling, which was
 interpreted as a strong coupling IR fixed point.

  From the holographic principle and
  the AdS/CFT correspondence,
 this renormalization group flow of type 0 string theory,
 can be understood to correspond to moving the brane a finite
 distance in the bulk. Motivated by this behaviour of
 type 0 theory, we will discuss in this talk
 the cosmological evolution of the brane-universe as the brane
 moves from the UV to the IR fixed point.

We calculate the effective energy density which is induced
 on the brane because of its motion in the particular background
 of a type 0 string. Using the approximate solutions of
 {\cite{Tset,Minah}, we find that for large values of the
 radial coordinate $r$, in the UV region, the effective energy density
 takes a constant value, which means that the universe has an
 inflationary period.

 \section{Brane-Universe}

 In \cite{Keh} it was  considered a D-brane moving in a generic
 static, spherically symmetric background. As the brane moves in a
 geodesic, the induced world-volume metric becomes a function of
 time, so there is a cosmological evolution from the brane point
 of view. The metric of a D3-brane is parametrized as
\begin{equation}\label{in.met}
ds^{2}_{10}=g_{00}(r)dt^{2}+g(r)(d\vec{x})^{2}+
  g_{rr}(r)dr^{2}+g_{S}(r)d\Omega_{5}
\end{equation}
 and there is also a dilaton field $\Phi$ as well as a $RR$
 background~$C(r)=C_{0...3}(r)$ with a self-dual field strength. The
 action on the brane is given by
\begin{eqnarray}\label{B.I. action}
  S&=&T_{3}~\int~d^{4}\xi
  e^{-\Phi}\sqrt{-det(\hat{G}_{\alpha\beta}+(2\pi\alpha')F_{\alpha\beta}-
  B_{\alpha\beta})}
   \nonumber \\&&
  +T_{3}~\int~d^{4}\xi\hat{C}_{4}+anomaly~terms
\end{eqnarray}
 The induced metric on the brane is
\begin{equation}\label{ind.metric}
  \hat{G}_{\alpha\beta}=G_{\mu\nu}\frac{\partial x^{\mu}\partial x^{\nu}}
  {\partial\xi^{\alpha}\partial\xi^{\beta}}
\end{equation}
 with similar expressions for $F_{\alpha\beta}$ and
 $B_{\alpha\beta}$.In the static gauge,
 using (\ref{ind.metric}) we can calculate the bosonic part of the
 brane Lagrangian which reads
\begin{equation}\label{brane Lagr}
L=\sqrt{A(r)-B(r)\dot{r}^{2}-D(r)h_{ij}\dot{\varphi}^{i}\dot{\varphi}^{j}}
-C(r)
\end{equation}
where $h_{ij}d \varphi ^{i} d \varphi^{j}$ is the line
 element of the unit five-sphere, and
\begin{equation}\label{met.fun}
  A(r)=g^{3}(r)|g_{00}(r)|e^{-2\Phi},
  B(r)=g^{3}(r)g_{rr}(r)e^{-2\Phi},
  D(r)=g^{3}(r)g_{S}(r)e^{-2\Phi}
\end{equation}
 Demanding conservation of energy $E$ and of total angular
 momentum $ \ell ^{2} $ on the brane, the induced four-dimensional metric
 on the brane is
\begin{equation}\label{fin.ind.metric}
d\hat{s}^{2}=-d\eta^{2}+g(r(\eta))(d\vec{x})^{2}
\end{equation}
 with $\eta$ the cosmic time which is defined  by
\begin{equation}\label{cosmic}
 d\eta=\frac{|g_{00}|g^{\frac{3}{2}}e^{-\Phi}}{|C+E|}dt
\end{equation}

 This equation is the standard form of a flat expanding universe.
If we define the scale factor as $\alpha^{2}=g$ then we can
calculate the Hubble constant $H=\frac{\dot{\alpha}}{\alpha}$,
where dot stands for derivative with respect to cosmic time. Then
we can interpret the quantity $(\frac{\dot{\alpha}}{\alpha})^{2}$
as an effective matter density on the brane with the result
\begin{equation}\label{dens}
\frac{8\pi}{3}\rho_{eff}=\frac{(C+E)^{2}g_{S}e^{2\Phi}-|g_{00}|(g_{S}g^{3}+\ell^{2}e^{2\Phi})}
{4|g_{00}|g_{rr}g_{S}g^{3}}(\frac{g'}{g})^{2}
\end{equation}

Therefore the motion of a D3-brane on a general spherically
symmetric background had induced on the brane a matter density. As
it is obvious from the above relation, the specific form of the
background will determine the cosmological evolution on the brane.

\section{Type 0 string background}

 The action of the type 0 string is given
by \cite{Tset}
\begin{eqnarray}\label{action}
S_{10}&=&~\int~d^{10}x\sqrt{-g}\Big{[} e^{-2\Phi} \Big{(}
 R+4(\partial_{\mu}\Phi)^{2} -\frac{1}{4}(\partial_{\mu}T)^{2}
-\frac{1}{4}m^{2}T^{2}-\frac{1}{12}H_{\mu\nu\rho}H^{\mu\nu\rho}\Big{)}
\nonumber \\&& - \frac{1}{4}(1+T+\frac{T^{2}}{2})|F_{5}|^{2}
\Big{]}
\end{eqnarray}
The tachyon is coupled to the $RR$ field through the function
\begin{equation}\label{ftac}
 f(T)=1+T+\frac{1}{2} T^{2}
\end{equation}
In the background where the tachyon field acquires vacuum
expectation value $T_{vac}=T_{0}=-1$, the tachyon function
(\ref{ftac}) takes the value $f(T_{0})=\frac{1}{2}$ which
guarantee the stability of the theory \cite{Tset}.

Technically it is easier to solve the equations of motion
resulting from the above action, if we go to new variables. One
can then introduce new parameter $ \rho $ through the relation
\begin{equation}\label{rho}
\rho = \frac{e^{2\Phi_{0}}}{4r^{4}}
\end{equation}
and the fields $ \xi $ and $\eta$ from the relations
\begin{equation}\label{xxi}
g = e^{\frac{\Phi-\xi}{2}} , g_{s} = e^{\frac{\Phi+\xi}{2}-\eta}
\end{equation}
Then (\ref{in.met}) takes the form,
\begin{equation}
ds^{2} = - e^{\frac{\Phi-\xi}{2}}dt^{2}+
e^{\frac{\Phi-\xi}{2}}d\vec{x}^{2}+e^{\frac{\Phi+\xi}{2}-5\eta}d\rho^{2}
 +e^{\frac{\Phi+\xi}{2}-\eta}d\Omega_{5}^{2}
\end{equation}

 With this form of the metric, the action (\ref{action}) can be described by the
following Toda-like mechanical system (dot denotes
$\rho$-derivative)
\begin{equation}\label{Tota}
S = \int d\rho [\frac{1}{2}\dot{\Phi}^{2}
+\frac{1}{2}\dot{\xi}^{2} +\frac{\dot{T}^{2}}{4} -5\dot{\eta}^{2}
-V(\Phi,\xi,T,\eta)]
\end{equation}
with the potential $ V(\Phi,\xi,T,\eta)$ given by
\begin{equation}\label{Poten}
V(\Phi,\xi,T,\eta) = g(T)e^{\frac{1}{2}\Phi + \frac{1}{2}\xi
-5\eta} +20 e^{-4\eta} -Q^{2}f^{-1}(T)e^{-2\xi}
\end{equation}

 If the tachyon field takes its vacuum value and the dilaton
field a constant value $\Phi=\Phi_{0}$ one can find the
electrically charged three-brane
\begin{equation}\label{Sol}
  g_{00}=-H^{-\frac{1}{2}},
  g(r)=H^{-\frac{1}{2}},  g_{S}(r)=H^{\frac{1}{2}}r^{2},
g_{rr}(r)=H^{\frac{1}{2}},    H=1+\frac{e^{\Phi_{0}}Q}{2r^{4}}
\end{equation}
if the following ansatz for the $RR$ field is used
\begin{equation}\label{Form}
  C_{0123}=A(r),   F_{0123r}=A'(r)
\end{equation}

If $T$ and $\Phi$ are functions of the coordinate $r$, then
approximate solutions exist \cite{Tset,Minah}. These solutions are
valid for large (UV) and small (IR) values of the radial
coordinate. From the action (\ref{Tota}) we can derive the
following equations of motion \cite{Minah}
\begin{equation}\label{eqxi}
\ddot{\xi} +\frac{1}{2}g(T)e^{\frac{1}{2}\Phi + \frac{1}{2}\xi
-5\eta} +2\frac{Q^{2}}{f(T)}e^{-2\xi}=0
\end{equation}
\begin{equation}\label{eqeta}
\ddot{\eta} + \frac{1}{2}g(T)e^{\frac{1}{2}\Phi + \frac{1}{2}\xi
-5\eta} +8e^{-4\eta} =0
\end{equation}
\begin{equation}\label{eqphi}
\ddot{\Phi}+\frac{1}{2}g(T)e^{\frac{1}{2}\Phi + \frac{1}{2}\xi
-5\eta} =0
\end{equation}
\begin{equation}\label{eqT}
\ddot{T} +2g'(T)e^{\frac{1}{2}\Phi + \frac{1}{2}\xi -5\eta}
+2\frac{Q^{2}f'(T)}{f^{2}(T)}e^{-2\xi} = 0
\end{equation}
where $g(T)$ is the bare tachyon potential,
\begin{equation}\label{gfun}
g(T)=\frac{1}{2} T^{2}- \lambda T^{4}
\end{equation}
and $\lambda$ is a parameter. Defining a new variable
$\rho=u^{-4}$, in the UV for $u\longrightarrow\infty$, or
$\rho\longrightarrow 0$,
 we can solve the equations of motion
(\ref{eqxi})-(\ref{eqT}) to the next to leading order and  find
\cite{Minah,Tset}
\begin{equation}\label{apprT}
T = T_{0}-4\frac{g'(T_{0})}{g(T_{0})}\frac{1}{log \rho} +
O(\frac{log(-log\rho)}{log^{2}\rho})
\end{equation}
\begin{equation}\label{apprphi}
\Phi = -2log(C_{0}log\rho) -
\Big{(}7+8(\frac{g'(T_{0})}{g(T_{0})})^{2}\Big{)}
\Big{(}\frac{log(-log\rho)}{log\rho}\Big{)} +\frac{B}{log\rho} +
O(\frac{log^{2}(-log\rho)}{log^{2}\rho})
\end{equation}
\begin{equation}\label{apprxi}
\xi = log\Big{(}\sqrt{2f^{-1}(T_{0})}Q\rho\Big{)} -
\frac{1}{log\rho} + O(\frac{log(-log\rho)}{log^{2}\rho})
\end{equation}
\begin{equation}\label{appreta}
\eta = \frac{1}{2}log(4\rho)- \frac{1}{log\rho} +
O(\frac{log(-log\rho)}{log^{2}\rho})
\end{equation}
 where $C_{0}=- \frac{4C_{2}^{5}} {g(T_{0}) \surd C_{1}} $ and
\begin{equation}\label{Cfun}
C_{1} =
\frac{2Q}{\sqrt{2f(T_{0})}}(1+\frac{1}{4log\frac{u}{u_{0}}}) ,
C_{2} = 2(1+\frac{1}{4log\frac{u}{u_{0}}})
\end{equation}

The above solutions show that at the UV point the tachyon takes a
constant value and if we calculate the effective coupling
$e^{\frac{1}{2}\Phi}$ we will see that it goes to zero for large
$u$.

For $u\longrightarrow 0$ or for large $\rho$ the approximate
solutions are
\begin{equation}\label{tset1}
T = -\frac{16}{log\rho}- \frac{8}{log^{2}\rho}(9loglog\rho-3) +
O(\frac{log^{2}log\rho}{log^{3}\rho})
\end{equation}
\begin{equation}\label{tset2}
\Phi = -\frac{1}{2}log(2Q^{2}) +2loglog\rho -
\frac{1}{log\rho}9loglog\rho + O(\frac{loglog\rho}{log^{2}\rho})
\end{equation}
\begin{equation}\label{tset3}
\xi = \frac{1}{2}log(2Q^{2}) + log\rho + \frac{9}{log\rho}
+\frac{9}{2log^{2}\rho}(9loglog\rho-\frac{20}{9}) +
O(\frac{log^{2}log\rho}{log^{3}\rho})
\end{equation}
\begin{equation}\label{tset4}
\eta = log2 + \frac{1}{2}log\rho + \frac{1}{log\rho}
+\frac{1}{2log^{2}\rho}(9loglog\rho-2) +
O(\frac{log^{2}log\rho}{log^{3}\rho})
\end{equation}

The tachyon field in the IR point goes to zero while the effective
coupling gets infinite. It is important to observe that the
approximate solutions in the UV (\ref{apprT})-(\ref{appreta}) and
in the IR (\ref{tset1})-(\ref{tset4}) of \cite{Tset} are related
by $ y \longrightarrow -y$, suggesting that they can be smoothly
connected into a full interpolating solution. As we mention above,
the tachyon field starts at T=-1 at $\rho=0$ in the UV , and grows
according to (\ref{apprT}), then enters an oscillating regime and
finally relaxes to zero according to (\ref{tset1}), when
$\rho=\infty$ in IR. This guarantees that
$T^{2}e^{\frac{1}{2}\Phi}$ becomes small which leads the metric in
the $AdS_{5} \times S^{5}$ form.

There is a question if we can trust the asymptotic solutions in
the infrared. The problem is that when the coupling becomes
strong, string corrections become important. The situation is
different in the UV where we can trust our solutions because the
coupling is small. The role of the $\alpha^{'}$ corrections in the
IR has been discussed in \cite{Minah}. It was claimed that the
$\alpha^{'}$ corrections are small.

\section{Cosmological evolution of the Brane-Universe }

We consider a D3-brane moving along a geodesic in the background
of a type 0 string. The metric on the brane (\ref{in.met}) using
the background solution (\ref{Sol}) is
\begin{equation}\label{Ind.Sol}
 d\hat{s}^{2}=(-H^{-\frac{1}{2}}+H^{\frac{1}{2}}\dot{r}^{2}
 +H^{\frac{1}{2}}r^{2}h_{ij}\dot{\varphi}^{i}\dot{\varphi}^{j})dt^{2}
 +H^{-\frac{1}{2}}(d\vec{x})^{2}
\end{equation}
The $RR$ field $C=C_{0123}$ using the ansatz (\ref{Form}) becomes
\begin{equation}\label{cbar}
  C^{~'}=2 Q g^{2}g^ {-\frac{5}{2}}_{s}\sqrt{g_{rr}}f^{-1}(T)
\end{equation}
where Q is a constant. Using again the solution (\ref{Sol}) the
$RR$ field can be integrated to give
\begin{equation}\label{Cterm}
C=e^{-\Phi_{0}}f^{-1}(T)(1+\frac{e^{\Phi_{0}}Q}{2r^{4}})^{-1}+Q_{1}
\end{equation}
where $Q_{1}$ is a constant. The effective density on the brane
(\ref{dens}), using eq.(\ref{Sol}) and (\ref{cbar})  becomes
\begin{equation}\label{cre}
\frac{8\pi}{3}\rho_{eff}=\frac{1}{4}[(f^{-1}(T)+EHe^{\Phi_{0}})^{2}-(1+\frac{\ell^{2}e^{
2\Phi_{0}}}{2}H)]
\frac{Q^{2}e^{2\Phi_{0}}}{r^{10}}H^{-\frac{5}{2}}
\end{equation}
where the constant $Q_{1}$ was absorbed in a redefinition of the
parameter $E$. Identifying $g=\alpha^{2}$ and using
$g=H^{-\frac{1}{2}}$ we get from (\ref{cre})
\begin{eqnarray}\label{aro}
\frac{8\pi}{3}\rho_{eff}&=&(\frac{2e^{-\Phi_{0}}}{Q})^{\frac{1}{2}}
 \Big{[} \Big{(} f^{-1}(T)+\frac{Ee^{\Phi_{0}}}{\alpha^{4}} \Big{)}^{2}
-\Big{(}1+\frac{\ell^{2}e^{2\Phi_{0}}}
{\alpha^{6}}(\frac{2e^{-\Phi_{0}}}{Q})^{\frac{1}{2}}\nonumber \\&&
(1-\alpha^{4})^{\frac{1}{2}} \Big{)}  \Big{]} (1-\alpha^{4})
^{\frac{5}{2}}
\end{eqnarray}
From the relation $g=H^{-\frac{1}{2}}$ we can see that the range
of $\alpha$ is $0\leq \alpha <1$, while the range of $r$ is $0\leq
r< \infty$. We can calculate the scalar curvature of the
four-dimensional universe as
\begin{equation}\label{curv}
  R_{brane}=8\pi(4+\alpha\partial_{\alpha})\rho_{eff}
\end{equation}
If we use the effective density of (\ref{aro}) it is easy to see
that $R_{brane}$ of (\ref{curv}) blows up at $\alpha=0$. On the
contrary if $r\rightarrow 0$, then the $ds^{2}$ of (\ref{in.met})
becomes
\begin{equation}\label{ads}
ds^{2}_{10}= \frac{r^{2}}{L} (-dt^{2}+(d\vec{x})^{2})+
      \frac{L}{r^{2}} dr^{2}+  L d\Omega_{5}
\end{equation}
with $L=(\frac{e^{\Phi_{0}}Q}{2})^{\frac{1}{2}}$. This space  is a
regular $AdS \times S^{5}$ space.

Therefore the brane develops an initial singularity as it reaches
$r=0$, which is a coordinate singularity and otherwise a regular
point of the $AdS_{5}$ space. This is another example in Mirage
Cosmology \cite{Keh} where we can understand the initial
singularity as the point where the description of our theory
breaks down.

If we take $\ell^{2}=0$, set the function $f(T)$ to its minimum
value and also taking $\Phi_{0}=0$, the effective density
(\ref{aro}) becomes
\begin{equation}\label{laro}
\frac{8\pi}{3}\rho_{eff}=(\frac{2}{Q})^{\frac{1}{2}}
\Big{(}(2+\frac{E}{\alpha^{4}})^{2} -1 \Big{)} (1-\alpha^{4})
^{\frac{5}{2}}
\end{equation}
As we can see in the above relation, there is a constant term,
coming from the tachyon function $f(T)$. For small $\alpha$ and
for some range of the parameters $E$ and $Q$ it gives an
inflationary phase to the brane cosmological evolution. In Fig.1
we have plotted $\rho_{eff}$ as a function of $\alpha$ for $Q=2$.

\begin{figure}[h]
\centering
\includegraphics[scale=0.7]{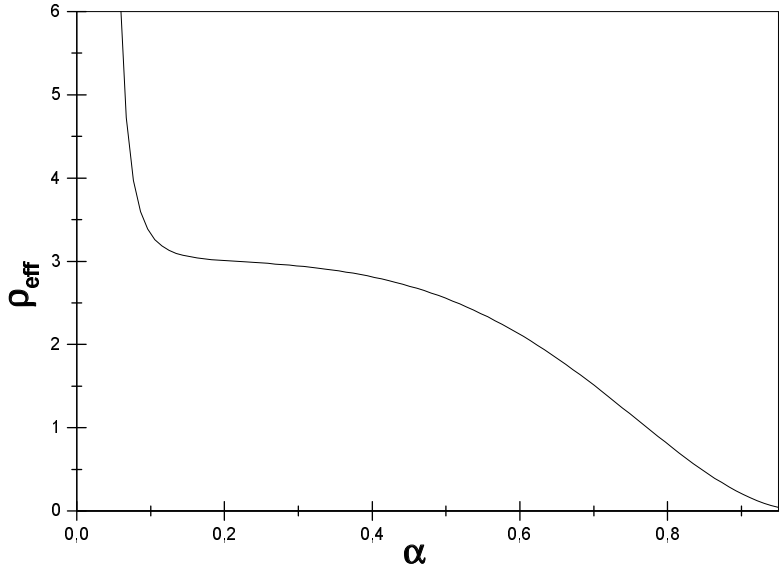}\caption
{The induced energy density on the brane as a function of the
brane scale factor.}
\end{figure}

As the brane moves away from $r=0$ to larger values of $r$, the
universe after the inflationary phase enters a radiation dominated
epoch because the term $\alpha^{-4}$ takes over in (\ref{laro}).
As the cosmic time $\eta$ elapses the $\alpha^{-8}$ term dominates
and finally when the brane is far away from $r=0$, the term which
is controlled by the angular momentum $\ell^{2}$ gives the main
contribution to the effective density. Non zero values of
$\ell^{2}$ will give negative values for $\rho_{eff}$. We expect
that at later cosmic times there will be other fields, like gauge
fields, which will give a different dynamics to the cosmological
evolution and eventually cancel the negative matter density.

Let us now see what happens if the tachyon and dilaton fields are
not constant. In the presence of a non trivial tachyon field the
coupling $e^{-\Phi}$ which appears in the Dirac-Born-Infeld action
in (\ref{B.I. action}), is modified by a tachyonic function
$\kappa(T)=1+\frac{1}{4}T+O(T^{2})$. Then we can define an
effective coupling \cite{Typ0}
\begin{equation}\label{effphi}
e^{-\Phi}_{eff}=\kappa(T) e^{-\Phi}
\end{equation}

 The bulk
fields are also coordinate dependent and the induced metric on the
brane  will depend on a non trivial way on the dilaton field.
Therefore the metric in the string frame will be connected to the
metric in the Einstein frame through
$g_{St}=e^{\frac{\Phi}{2}}_{eff} g_{E}$.  All the quantities used
so far were defined in the string frame. We will follow our
cosmological evolution in the Einstein frame. Then the relation
(\ref{dens}) becomes
\begin{equation}\label{eindens}
 \frac{8\pi}{3}\rho_{eff}= (\frac {\dot{\alpha}}{\alpha})^{2}=
\frac{(C+E)^{2}g_{S}-|g_{00}|(g_{S}g^{3}+\ell^{2})}
{4|g_{00}|g_{rr}g_{S}g^{3}}(\frac{g'}{g})^{2}
\end{equation}

Having the approximate solution in the UV given by
(\ref{apprT})-(\ref{appreta}) we can calculate the metric
components of the metric (\ref{in.met}) and find
\begin{equation}\label{gyy}
g_{yy} = \frac{1}{16}\sqrt{\frac{Q}{2}}(1-\frac{9}{2y}) , g_{s} =
\sqrt{\frac{Q}{2}}(1-\frac{1}{2y})
\end{equation}
\begin{equation}\label{g}
g = \frac{1}{\sqrt{2Q}}e^{\frac{y}{2}}(1-\frac{1}{2y})
\end{equation}
The variable y is defined by $\rho = e^{-y}$
 Then we can identify $g$ of (\ref{g}) with the scale factor $\alpha^{2}$
 and solve for $y$. We get two solutions
\begin{equation}\label{y1}
y_{1} = -\frac{1}{4log\alpha + log2Q},y_{2} = log2Q + 4log\alpha +
\frac{1}{log2Q + 4log\alpha}
\end{equation}
From the solution $y_{2}$ which has the right behaviour for large
$\alpha$, we keep the $log2Q + 4log\alpha$ term. Then the $RR$
field C becomes
\begin{equation}\label{cfield}
C = \frac{e^{y}}{Q} - \frac{2}{Q}Ei[y]
\end{equation}

Then, we can calculate the effective energy density from
(\ref{eindens}) setting $\ell^{2}=0$ and we get

\begin{eqnarray}\label{rhonimah}
\frac{8\pi}{3}\rho_{eff}& =&
 \Big{[} \Big{(}1-\frac{1}{Q\alpha^{4}}Ei[log2Q + 4log\alpha] +
\frac{E} {2\alpha^{4}} \Big{)}^{2} -
\frac{1}{4}\Big{(}1-\frac{1}{2(log2Q +
4log\alpha)}\Big{)}^{4}\nonumber \\&& \Big{
]}\Big{(}1-\frac{1}{2(log2Q +
4log\alpha)}\Big{)}^{-4}\Big{(}1-\frac{9}{2(log2Q +
4log\alpha)}\Big{)}^{-1} \nonumber \\&& \Big{(}1+\frac{1}{(log2Q +
4log\alpha)^{2}}\frac{1}{\Big{(}1-\frac{1}{2(log2Q +
4log\alpha)}\Big{)}}\Big{)}^{2}
\end{eqnarray}

For some typical value of the parameters Q=1 and E=1, and for
large values of $\alpha$, it is obvious that $\rho_{eff}$ has a
constant value. Therefore an observer on the brane will see an
expanding inflating universe. It is interesting to see what
happens for small values of $\alpha$. As $\alpha$ gets smaller, a
term proportional to $\frac{1}{(\log\alpha)^{4}}$ starts to
contribute to $\rho_{eff}$. Therefore the universe for small
values of scale factor has a slow expanding inflationary phase.
For smaller value of $\alpha$ we cannot trust the solution which
is reflected in the fact that $\rho_{eff}$ gets infinite. The
behaviour of the effective energy density as a function of the
scale factor is shown in Fig.2

\begin{figure}[h]
\centering
\includegraphics[scale=0.7]{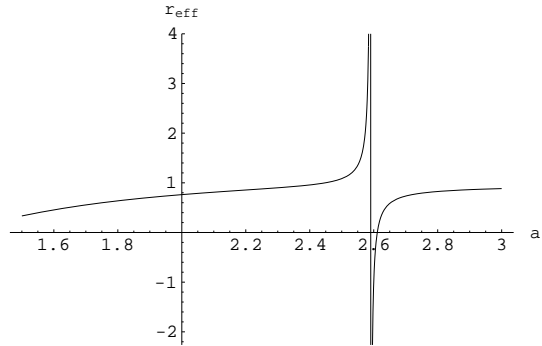}\caption
{The induced energy density on the brane as a function of the
brane scale factor.}
\end{figure}


 Going now to IR using (\ref{tset1})-(\ref{tset4}) we get for
the metric components
\begin{equation}\label{tsegyy}
g_{yy} = \frac{\sqrt{Q}}{16}2^{-\frac{3}{4}}(1-\frac{1}{2y}) ,
g_{s} = 2^{-\frac{3}{4}} \sqrt{Q}(1+\frac{7}{2y})
\end{equation}
\begin{equation}\label{tseg}
g =
\frac{2^{-\frac{1}{4}}}{\sqrt{Q}}e^{-\frac{y}{2}}(1-\frac{9}{2y})
\end{equation}
where now y is defined by $\rho = e^{y}$.

Then the identification $g=\alpha^{2}$ using (\ref{tseg}) gives
again two solutions
\begin{equation}\label{tsey1}
y_{1} = -\frac{9}{4log\alpha + log\sqrt{2}Q},y_{2} = -log\sqrt{2}Q
- 4log\alpha + \frac{9}{log\sqrt{2}Q + 4log\alpha}
\end{equation}
 For small $\alpha$ we keep from the solution $y_{2}$
 the term $ -log\sqrt{2}Q - 4log\alpha$. Using this solution we can
 calculate the $RR$ field
\begin{equation}\label{ctset}
C =- \frac{e^{-y}}{Q} - \frac{2}{Q}Ei[-y]
\end{equation}
Then $\rho_{eff}$ becomes,
\begin{eqnarray}\label{tserho}
\frac{8\pi}{3}\rho_{eff}& =& \Big{[} \Big{
(}-1-2\frac{1}{\sqrt{2}Q\alpha^{4}}Ei[log\sqrt{2}Q + 4log\alpha] +
\frac{E}{\sqrt{2}\alpha^{4}} \Big{)}^{2} - \nonumber
\\ &&
 \frac{1}{2}\Big{(}1+\frac{9}{2(log\sqrt{2}Q +
4log\alpha)}\Big{)}^{4} \Big{]}\Big{(}1+ \frac{9}{2(log\sqrt{2}Q +
4log\alpha)}\Big{)}^{-4} \nonumber \\&&
\Big{(}1+\frac{1}{2(log\sqrt{2}Q + 4log\alpha)}\Big{)}^{-1}
\nonumber
\\&&
\Big{(}1-\frac{9}{2(4log\alpha+ log\sqrt{2}Q)^{2}}
 \frac{1}{\Big{(}1-\frac{9}{2(log\sqrt{2}Q + 4log\alpha)}\Big{)}}\Big{)}^{2}
\end{eqnarray}

As we can see, the above relation is the same as the energy
density in the UV (relation (\ref{rhonimah})) up to some numerical
factors, as expected. The difference is, that now it is valid for
small $\alpha$. For small $\alpha$ first the term $
\frac{1}{\alpha^{8}}$ dominates and then the term
$\frac{1}{\alpha^{4}}$. As $\alpha$ increases the term $
\frac{1}{(log\alpha)^{4}}$ takes over and drives the universe to a
slow inflationary expansion.

\section{Discussion}

We had followed a probe brane along a geodesic in the background
of type 0 string. Assuming that the universe is described by a
three-dimensional brane, we calculate the effective energy density
which is induced on the brane because of this motion. We studied
this mirage matter as the brane-universe moves along the radial
coordinate.

Using an exact solution of the type 0 string theory, we found
\cite{Papa} that the motion of the brane-universe in this
particular background induces an inflationary phase on the brane.
In the case when the dilaton and tachyon fields are coordinate
dependent, there are approximate solutions for large values of the
radial coordinate, in the UV region and solutions for small values
of the radial coordinate in the IR. In the UV the effective
coupling of the theory is small, so we can trust the approximate
solutions. In the IR, the effective coupling becomes strong but it
was shown in the literature \cite{Minah}, that all string
corrections are small.

Using these solutions, we calculated the energy densities that are
induced on the brane. What we found is that for large values of
the scale factor as it is measured on the brane (large values of
the radial coordinate) the universe enters a slow inflationary
phase, in which the energy density is proportional to an inverse
power of the logarithm of the scale factor. As the scale factor
grows the induced energy density takes a constant value and the
universe enters a normal exponential expansion. For small values
of the scale factor the induced energy density scales as the
inverse powers of the scale factor and then the logarithmic terms
take over and the universe enters a slow exponential expansion.

\section*{Acknowledgement}
Work partially supported by the NTUA program Archimedes.

 \end{document}